\documentclass[aps,prd,draft,showkeys,floatfix]{revtex4}

\usepackage{graphicx}
\usepackage{amsmath,amssymb,latexsym}
\usepackage{bm}
\usepackage{epsfig}

\begin{document}

\begin{flushright}
\begin{obeylines}
UG-FT-215/07
CAFPE-85/07
\end{obeylines}
\end{flushright}
\vspace*{0.7cm}

\title{Spin Dependence of Heavy Quark Fragmentation}

\author{Fernando Cornet}
\affiliation{Departamento de F\'{\i}sica Te\'orica y del Cosmos and
Centro Anadaluz de F\'{\i}sica de Part\'{\i}culas, Universidad de
Granada, E-18071 Granada, Spain}

\author{Carlos A. Garc\'{\i}a Canal}
\affiliation{Departamento de F\'{\i}sica, Universidad Nacional de La
Plata, C.C.67, La Plata (1900), Argentina}

\begin{abstract}
\noindent
We propose that the non-perturbative fragmentation functions 
describing the transition
from a heavy quark to a heavy meson is proportional to the
square of the produced meson wave function at the origin. We
analyze the effects of this proposal on the number of pseudoscalar
mesons compared to the number of vector mesons produced and find a good
agreement with experimental data. Finally, we discuss further
experimental checks for our hypothesis.
\end{abstract}

\pacs{~}

\maketitle


Heavy quark production in high energy collisions, either 
$e^+e^-$, $ep$,
$pp$ or $p \overline{p}$, provides a good laboratory 
to test QCD in both, 
its perturbative and non-perturbative sectors. In all the experiments both 
sectors contribute. In processes with a hadron in the initial state one has to
consider the quark and gluon distribution 
functions in the initial hadron and
the heavy quark fragmentation function describing the transition of the
heavy quark into the measured final state hadronic system. In $e^+ e^-$
processes only this last non-perturbative piece contributes. It is, thus, 
clear that in order to get the maximum theoretical information from the 
increasingly precise experimental data on heavy quark production one 
must have a good description of all the pieces involved in the calculation, 
in particular of the fragmentation functions. 

The heavy quark fragmentation functions receive two contributions. 
In high
energy processes the heavy quark is produced far off mass-shell and it
emits gluons until it becomes on mass-shell. This process can be calculated
perturbatively via the usual DGLAP equations describing the evolution. In this 
way one obtains a better theoretical prediction for the partonic 
cross-section. At 
this point one still has to take into account the transition from the heavy 
quark into a hadron which is described by a purely non-perturbative 
fragmentation function. Since all the scaling violations are taken into 
account via the perturbative calculation, one expects the 
non-perturbative fragmentation functions to be scale independent 
\cite{mele}.

The experimental situation on charm and bottom production in different 
collision events has been reviewed in \cite{gladilin}. These
measurements are mainly devoted to test perturbative QCD and to
extract fragmentation functions and fractions. The best description of the
experimental data  is obtained by using either a one
free parameter non-perturbative phenomenological parametrization by
Kartvelishvili, Likhodede and Petrov \cite{K} or a two parameter one
by Bowler \cite{bowler}. The parametrizations by Peterson,
Schlatter, Schmitt and Zerwas \cite{pete} is also widely used when analyzing
experimental data. Other popular parametrizations 
are: Collins and Spiller
\cite{collins}, Colangelo and Nason \cite{cola} and Braaten, Cheung,
Fleming and Yuan \cite{braaten}. 

All these parametrizations provide different realizations of the
original Bjorken \cite{bjorken} and Suzuki \cite{suzuki} proposal 
that the heavy quark fragmentation
function, contrary to what happens with light quarks, should be very hard,
i.e. the heavy quark should retain most of its momentum in the hadronization
process. The exact shape of the dependence of the fragmentation function
on the heavy quark momentum is controlled, in the 
parametrizations cited above,
by some free parameters (the number of free parameters  depends on the 
parametrization) that have to be fiited to the experimental data. In 
general these parameters have no absolute physical meaning
(see for example \cite{yao} for a discussion). 

In addition to the dependence of the fragmentation functions with the 
heavy quark momentum, the different experiments have also measured the 
relative number of charmed and bottom mesons produced in the 
pseudoscalar state ($D$ and $B$) and in the vector 
state ($D^\star$, $B^\star$). More specifically, they have measured
the quantiy $P_V$ defined by
\begin{equation} \label{PV}
P_V = \frac{V}{P+V}
\end{equation}
where $V$ and $P$ stand for number of vector and pseudoscalar states
produced, respectively. The present experimental values for
charmed mesons are:
\begin{equation}
\label{datac}
\begin{array}{lcl}
P_V^D = 0.566 \pm 0.025^{+ 0.007+0.022}_{- 0.022-0.023} & \quad \quad \quad &
\hbox{ZEUS (photoproduction) \cite{zeusph}} \vspace*{0.3cm}\\ 
P_V^D = 0.590 \pm 0.037^{+ 0.022}_{-0.018}              & \quad \quad \quad &
\hbox{ZEUS (DIS) \cite{zeusDIS}} \vspace*{0.3cm}\\ 
P_V^D = 0.693 \pm 0.045 \pm 0.004 \pm 0.009             & \quad \quad \quad &
\hbox{H1 \cite{h1}} \vspace*{0.3cm}\\
P_V^D = 0.614 \pm 0.023                                 & \quad \quad \quad &
\hbox{$e^+\,e^-$ average \cite{ee}}
\end{array}
\end{equation}
These values provide a world average
\begin{equation}
\label{avg}
\bar{P}_V^D = 0.611  \pm 0.016 .
\end{equation}

For $B$ mesons there is only one measurement 
available \cite{gladilin} 
\begin{equation}
\label{pvb}
P_V^B = 0.75 \pm 0.04 ^{+ 0.023}_{- 0.025}.
\end{equation}

The values for the charmed mesons are clearly smaller than the 
widely used naive spin counting prediction $P_V = 0.75$, but this is not 
the case for the bottom mesons. So, whatever mechanism in the
fragmentation process is claimed to be responsible
for the decrease of $P_Vĉ$ with respect to the naive prediction should,
in a natural way, produce a much smaller effect for mesons containing
a $b$ quark than for mesons containing a $c$ quark. 

Inspired by the fact that positronium production cross-section
(where the calculations are fully under control because the theory, QED, is
perturbative)
is proportional to the square of the wave function at 
the origin \cite{martynenko},
in this paper we propose that the fragmentation function of a heavy
quark, $Q$, into a heavy meson $M$ should be proportional to the 
square of the meson $M$ wave function at the origin.  
At lowest order in Heavy
Quark Effective Theory the wave functions at the origin for the 
pseudoscalar and vector states are the same and this would reproduce
the naive spin counting result, but they differ at $O(1/m_Q)$
and, obviously, the effects on the $P_V$ predictions will be larger
for mesons containing a $c$ quark than for mesons containing a $b$ quark. 
We, thus, propose to modify the naive $P_V$ expression (\ref{PV}) and to
consider instead
\begin{equation} \label{PVC}
P_V = \left( \frac{P}{V} + 1\right) ^{-1} =
\left(\frac{1}{3}\,\frac{|\psi_{P} (0)|^2}{|\psi_{V} (0)|^2} +
1\right)^{-1} .
\end{equation}
Notice that our predictions for the different rates $P_V$ will be
parameter free (once the wave functions at the origin are known),
because we only modify the normalization of the fragmentation
functions leaving their dependence on the heavy quark momentum
unchanged.

It is important to notice that in the analysis of semileptonic
meson decays a sensible improvement of the theoretical results
when compared with experimental data, was obtained by breaking heavy
quark symmetry and taking hyperfine interactions into account
\cite{scora}.

In the update \cite{scora} of the
Isgur-Scora-Grinstein-Wise model \cite{ISGW} one finds approximate
variational wave functions that consider separately each spin state.
The distinction between spin states is mandatory in order to get
agreement with the experimental data for the decays. The wave functions
for the lowest lying pseudoscalar and vector states
are written in terms of a single parameter $\beta_S$ as
\begin{equation}
\label{wf}
\psi_{1S} =
\frac{\beta_S^{3/2}}{\pi^{3/4}}\,\exp{\left(-\frac{1}{2}\,
                                             \beta_S^2\,r^2\right)}
\end{equation}
The value of $\beta_S$ is fixed for each meson to properly describe
its decay and the values obtained in Ref. \cite{scora} for different 
mesons are shown in Table I.

The results we obtain plugging the wave functions at the origin from
\cite{scora} into our Eq. (\ref{PVC}) are shown in the last
column of Table I. We obtain a sensible reduction in
the value of $P_V^D$
with respect to the naive
spin counting prediction and our result is in good agreement
with the experimental data. For $P_V^B$ the obtained reduction is much smaller
and the result is within one standard deviation of the experimental data,
Eq. (\ref{pvb}).
We should stress here once more that these numbers have been obtained 
without using any free parameter since the value of the $\beta_S$ 
parameter has been fixed in an independent analysis.

We have also calculated our predictions for the relative number of $D_s$
and $D_s^\star$ as well as $B_s$ and $B_s^\star$ and shown the results
in the last column of 
in Table I. The reduction with respect to the naive prediction is larger in
the mesons containing a heavy and a strange quark than in the mesons 
containing a heavy and a $u$ or $d$ quark. This is a prediction of
our assumption that can be experimentally checked. In particular, the 
low value of $P_V^{D_s}$ looks promising for such a test.

\begin{table}
\begin{tabular}{c|c|c|c|c}
\vspace{1mm}
{\underline{Meson}} & {\underline{Mass (MeV)}} & {\underline{$\beta_{S}$ (GeV)}} & {\underline{$P_V$}}  \\
$D$ & 1864.1 & 0.45 & &\\
 & & & 0.64 \\
$D^{\ast}$ & 2006.7 & 0.38 & & \\
\\
$D_s$ & 1969.0 & 0.56 & &\\
& &  & 0.59 \\
$D^{\ast}_s$ & 2112.0 & 0.44 & &\\
\\
$B$ & 5279.3 & 0.43 & & \\
& & & 0.71\\
$B^{\ast}$ & 5325.0 & 0.40 & & \\
\\
$B_s$& 5369.6 & 0.54 & &\\
& & & 0.69 \\
$B^{\ast}_s$ & 5412.8 & 0.49 & & \\
\end{tabular}
\caption{Values of the parameter $\beta_S$ entering in the
wave functions of the mesons, see Eq. (\ref{wf}) and predictions for 
heavy meson production rates}
\end{table}

The expression in Eq. (\ref{PVC}) is only valid if both the 
pseudoscalar and vector mesons entering have the same quark content. 
If only one of them contains a strange quark (in addition to the
heavy quark) one has to
take into account also the strange suppression factor that is usually
defined as the ratio of the number of charmed strange mesons 
with respect to
the number of charmed non-strange mesons 
produced. However, under our hypothesis one should be careful
because 
these ratios depend now on the wave functions at the origin. One 
would expect, however, that the strange suppression factor would only contain
information about the relative probability of producing 
a $s \bar{s}$ from the vacuum
compared to the probability of producing a pair of light quarks. This means,
one would expect the stange suppression factor to be independent of the 
spin of the produced mesons. So, in order 
to define such a spin independent strange suppression factor one has to use:
\begin{equation}
\label{ssf}
\gamma_{SI} = \frac{|\psi_D(0)|^2}{|\psi_{D_s}(0)|^2} \gamma_P =
         \frac{|\psi_{D^*}(0)|^2}{|\psi_{D_s^*}(0)|^2} \gamma_V.
\end{equation}
where $\gamma_P = \frac{D_s}{D}$ and $\gamma_V = \frac{D^\star_s}{D^\star}$
are the suppression factors measured in the pseudoscalar and vector channels,
respectively. Using the $\beta_S$ parameters listed in Table I, it is clear
that $\gamma_{SI} = 0.52 \gamma_P = 0.64 \gamma_V$.

Before ending we would like to briefly comment on the use of the
model for fragmentation functions proposed in Ref. \cite{braaten}
in computing the ratio $P_V$. In this case one has spin dependent
parametrizations depending on an extra parameter called $r$ 
related to the heavy quark and heavy meson masses. Clearly, in this
case one can obtain agreement with experimental data by looking for
the appropriate value of $r$, but also there is a dependence
on the square of the wave functions at the origin 
(see Eqs. (31) and (32) in \cite{braaten}) that, in order to be consistent
with the results in Ref. \cite{scora}, take different values for different 
mesons.

\section*{Final Remarks}
The naive spin counting prediction for the ratio $P_V$ does not fit
the experimental data, in particular in the case of charmed $D$
mesons. We propose that the fragmentation functions should be
proportional to the square of the produced meson wave function
at the origin. We have analyzed the effects of this proposal on the
values of $P_V$ for charmed and bottom mesons and found very good
agreement with the experimental data
using values for the wave functions at the origin fixed from meson
decays. As a way to check our proposal we estimate the
values of $P_V$ for charmed-strange and bottom-strange mesons for which
no experimental data are yet available. Further checks can be performed
measuring our proposed spin dependence of the strange suppession factor.

\section*{Acknowledgements}
We thank F. del Aguila, L. Labarga and M. Zambrana for useful discussions.
F.C. acknowledges financial support from Junta de Andalucia
(FQM-330) and Ministerio Espa\~nol de Educaci\'on y Ciencia
(FPA2006-05294). CAGC warmly acknowledges the hospitality
of Departamento de F{\'\i}sica Te\'orica y del Cosmos of
the Universidad de Granada that made possible this collaboration
and to the Junta de Andalucia for financial support during
his visit to Granada.

\end{document}